# Geodesic Deviation in the AdS Black String Spacetime

H.Culetu[*]

Ovidius University, Department of Physics,

B-dul Mamaia 124, 8700 Constanta, Romania

**Abstract.** The equation of motion of test particles in the geometry of a black string embedded in a five dimensional AdS spacetime is studied. With an inversion transformation on the fifth coordinate "w", we obtain a simple form of the geodesics and the geodesics deviation. The equations for the separation vector $\xi^\alpha$ between two nearby geodesics are found, with $\xi^5(\lambda) = \xi^5_{(0)} \sin(\lambda/l)$ independent of the Schwarzschild potential U(r) (l – the AdS radius).

## 1. Introduction

The conjecture that our world would be a three-brane embedded in extra spatial dimensions has been considered as a solution to the weak scale hierarchy problem.

The recent developments are based on the idea that ordinary matter fields are confined to a three – dimensional world while gravity could live in a higher dimensional space [1,2]. But the question is whether the model gives back standard four dimensional gravity in the brane assumed to represent our world.

Randall and Sundrum (RS) [3] consider our Universe as a negative tension domain wall separated from a positive tension wall by a region of anti - de Sitter geometry. In a 2–nd work [4] RS have introduced a positive tension 3-brane (where we live) inside 5-dimensional anti-de Sitter (AdS) space.

Chamblin et al. [5] studied a black hole formed by gravitational collapse in the RS models with the horizon extended into the directions transverse to the brane. To be in accordance with General Relativity (and therefore Birkhoff's theorem), the induced metric on the brane (domain wall) is Schwarzschild's and the black hole on the brane is in fact a black string in the 5 – dimensional spacetime.

*e-mail : hculetu@yahoo.com

In spite of the fact that the black string is unstable near the AdS horizon (the Gregory-Laflamme instability), it would be considered, however, stable far from the horizon, where the solution lookslike a "black cigar"[5]. Although the precise form of the "black cigar" metric (which describes a nonrotating, uncharged black hole) is not known, a black string well approximates a "black cigar" near the brane and in the case the black hole length scale is large enough relative to the bulk AdS length scale.

In the present article we examine in section 2 the geodesics of a test particle in the five – dimensional black string space of Chamblin et al. but with the fifth dimension undergoing an inversion. It is equivalent, as Grojean [6] pointed out, with a $Z_2$-symmetry transformation of the RS fifth coordinate y. The geodesics have a simpler form compared to that of Chamblin et al., with a harmonic oscillator type equation for the fifth coordinate (timelike geodesics) and a straightline (null geodesics).

In section 3 we study the geodesic deviation phenomenon taking into account two nearby geodesics with $\theta = \pi/2$ and $\varphi$ =const. (the angular momentum L=0). We found the fifth component of the spacelike separation vector $\xi^\alpha$ oscillates with the frequency $l^{-1}$, exactly as the coordinate itself. Concerning the time and radial components, more complicate equations were obtained. They have a similar form, but if U(r) were a linear function of the radial coordinate (which is valid near the horizon r=2m), both equations would acquire the very simple form of the equation for $\xi^5$.

## 2. The geodesic motion

The five dimensional AdS metric is given by

$$ds^2 = e^{-2y/l}\eta_{ij} dx^i dx^j + dy^2 , \tag{2.1}$$

where $\eta_{ij}$ = diag(-1, 1, 1, 1) and y is the fifth coordinate. The latin indices run from 0 to 3 and "l" is the AdS radius of curvature.

Chamblin et al.[5] introduce the coordinate z = lexp(y/l) to get a conformally flat geometry

$$ds^2 = \frac{l^2}{z^2}\left(dz^2 + \eta_{ij} dx^i dx^j\right). \tag{2.2}$$

They note that (2.2) still satisfies the Einstein equations provided the Minkowski metric $\eta_{ij}$ is replaced by a Ricci – flat one. Therefore, a black string on a three-brane (z=const.) embedded in a 5 – dimensional AdS space may be described by the metric

$$ds^2 = \frac{l^2}{z^2}\left(-U(r) dt^2 + \frac{dr^2}{U(r)} + r^2 d\Omega^2 + dz^2\right), \tag{2.3}$$

with $d\Omega^2 = d\theta^2 + \sin^2\theta \, d\varphi^2$ and $U(r) = 1 - 2m/r$. It is an easy task to include a $Z_2$ reflexion-symmetric domain wall (z = const.) : the junction conditions are satisfied provided the tension of the wall $\sigma = 6 / 8\pi G_5 l$.

We tried in this paper to obtain a simpler form of the geodesics equations using appropriate coordinates which express an exchange between the short and large distances (T duality [6]) (the distance to the brane is replaced by its inverse).

Let us change the 5-th coordinate z to :

$$w = \frac{l^2}{z}. \tag{2.4}$$

It is just a $Z_2$ symmetry operation in RS coordinates (to change the sign of y means to pass from z to w = l exp(-y/l) ). In terms of "w", the metric (2.3) becomes :

$$ds^2 = \frac{w^2}{l^2}\left(-U \, dt^2 + \frac{dr^2}{U} + r^2 \, d\Omega^2\right) + \frac{l^2}{w^2} dw^2, \tag{2.5}$$

The velocity along a timelike or null geodesic is denoted by $u^\alpha$ ($\alpha$ = 0, 1, 2, 3, 5, in the order t, r, $\theta$, $\varphi$, w). The independence of $g_{\mu\nu}$ on t and $\varphi$ gives two Killing vectors $k^\alpha_{(t)}$ and $k^\alpha_{(\varphi)}$ and two conserved quantities

$$E = -k^\alpha_{(t)} u_\alpha \quad , \quad L = k^\alpha_{(\varphi)} u_\alpha \, ,$$

with $k^\alpha_{(t)} = (1,0,0,0,0)$ and $k^\alpha_{(\varphi)} = (0,0,0,1,0)$. Therefore, the geodesic equations yield

$$\frac{dt}{d\lambda} = \frac{El^2}{Uw^2} \quad , \quad \frac{d\varphi}{d\lambda} = \frac{l^2}{w^2} \frac{L}{r^2}, \quad (\theta = \frac{\pi}{2}), \tag{2.6}$$

To get the equation for $w(\lambda)$, we have, from the general expression for the geodesic equations

$$\frac{d}{d\lambda}\left(\frac{l^2}{w^2}\frac{dw}{d\lambda}\right) + \frac{l^2 E^2}{U w^3} - \frac{w}{l^2 U}\left(\frac{dr}{d\lambda}\right)^2 - \frac{l^2 L^2}{r^2 w^3} + \frac{l^2}{w^3}\left(\frac{dw}{d\lambda}\right)^2 = 0, \tag{2.7}$$

where eqs.(2.6) have been used. The eq.(2.5) would read as

$$-1 = -\frac{l^2 E^2}{U w^2} + \frac{w^2}{l^2 U}\left(\frac{dr}{d\lambda}\right)^2 + \frac{l^2 L^2}{w^2 r^2} + \frac{l^2}{w^2}\left(\frac{dw}{d\lambda}\right)^2. \tag{2.8}$$

Combining the last two equations to get rid of the term with $dr/d\lambda$, one finds, for the timelike geodesics:

$$\frac{d^2 w}{d\lambda^2} + \frac{1}{l^2} w = 0. \quad \text{whence} \quad w(\lambda) = w_0 \sin(\lambda/l), \tag{2.9}$$

with $w_0$ – the amplitude of $w(\lambda)$. Therefore, w undergoes harmonic oscillations around the horizon w = 0, with the frequency $l^{-1}$ (a very large value provided l is few orders of magnitude above the Planck value). From the point of view of a brane world observer, the test particle which orbits the black string looks as a "halo" of dark matter [7] on the brane. This means that the orbiting "halo" will actually be composed of "clouds" which are "breathing", i.e. expanding and contracting

As far as the null geodesic is concerned, similar steps lead to

$$\frac{d^2w}{d\lambda^2}=0, \qquad (2.10)$$

whence $w(\lambda) = a\lambda+b$ (a, b – constants of integration). Hence, in spite of the fact that the 5 – dimensional spacetime is curved, the trajectory of a photon on the 5 – th direction is a straightline.

Let us study now the radial motion. Once $w(\lambda)$ is known, we can determine $r(\lambda)$ from eq.(2.5), with $\theta = \pi/2$ and by using eqs. (2.6). It would be useful to consider $w_0 = l$, a case in which the amplitude of oscillations equals the radius of the AdS spacetime (otherwise, we reach the same result, rescaling the coordinates with the constant factor $w_0/l$). With this choice, eq.(2.5) yields

$$-1=-\frac{E^2}{U\sin^2\frac{\lambda}{l}}+\frac{\sin^2\frac{\lambda}{l}}{U}\left(\frac{dr}{d\lambda}\right)^2+\frac{L^2}{r^2\sin^2\frac{\lambda}{l}}+\cot^2\frac{\lambda}{l}. \qquad (2.11)$$

After some rearengements and using the first eq.(2.6), (2.11) can be written as

$$\left(E\frac{dr^*}{dt}\right)^2+U(r)\left(1+\frac{L^2}{r^2}\right)=E^2. \qquad (2.12)$$

where the tortoise coordinate [8] $r^* = r + 2m \ln(r/2m -1)$ has been introduced. It is just the radial equation for a timelike geodesic in the Schwarzschild 4 – dimensional spacetime [8]. Compared to Chamblin et al. expression (their eq. (3.11) ), our radial geodesics of Schwarzschild type were obtained with no use of some unusual relation between the four and five dimensional affine parameters.

To obtain the differential equation for a null radial geodesic, we make use of the eq.(2.8), with zero on the l.h.s. Keeping in mind that $w(\lambda) = \lambda$ ($\alpha = 1$, in order to have a velocity $dw/d\lambda$ equal to unity and $\beta = 0$, by an appropriate choice of the origin of $\lambda$), the equation for $r(t)$ is the same as that for timelike geodesic

$$\left(E\frac{dr^*}{dt^*}\right)^2+U(r)\left(1+\frac{L^2}{r^2}\right)=E^2, \qquad (2.13)$$

with $t^*$ - a new coordinate time, given by

$$\frac{dt^*}{d\lambda}=\frac{l^2 E}{\lambda^2 U(r)}.$$

The result is not surprising as Chamblin et al.[5] reached the same property :the radial equation has the same form for timelike and null geodesics In addition, our new four – dimensional affine parameter is just the coordinate time.

## 3. The geodesic deviation

As Misner et al. have noticed [8], the equation of geodesic deviation summarizes the entire effect of geometry on matter. It is the analog of Lorentz force law from electromagnetism.

Let us consider geodesics with $L = 0$ and $\theta = \pi/2$. Using the coordinate transformation (2.4), the components of the tangent vector $u^\alpha$ to the geodesic can be written as

$$u^\alpha = \left( \frac{l^2 E}{w^2 U}, \frac{l^2}{w^2}\sqrt{E^2 - \frac{w_0^2 U}{l^2}}, 0, 0, -\frac{w}{l}\sqrt{\frac{w_0^2}{w^2} - 1} \right), \quad (3.1)$$

Here $w_0$ plays a similar role with $l^2/z_1$, where $z_1$ is a constant introduced by the authors of [5]. From eq.(3.9) of [5] it is clear that $z_1 = |z_{min}(\lambda)|$. But $|z_{min}| = l^2/|w_{max}|$. Since $w_{max}$ for a timelike geodesic is $w_0$, we may take $w_0 = l^2/z_1$.

Having found the components of $u^\alpha$, we are in a position to write down the differential equations for the components of the (spacelike) separation vector $\xi^\alpha$ by means of the geodesic deviation equation

$$\frac{d^2 \xi^\alpha}{d\lambda^2} = R^\alpha{}_{\mu\nu\kappa} u^\mu \xi^\nu u^\kappa, \quad (3.2)$$

where $R^\alpha{}_{\mu\nu\kappa} = \partial_\kappa \Gamma^\alpha_{\mu\nu} - ...$ is the Riemann tensor and all the parameters are evaluated in a local Lorentz frame. With constant angular variables, $\xi^\nu$ has only three nonzero components: $\xi^0$, $\xi^1$ and $\xi^5$. Therefore, the greek indices take here only values 0, 1, 5. The nonzero components of the Riemann tensor which are of interest to us, in the geometry (2.5) are the following

$$R^5{}_{050} = -\frac{w^2 U}{l^4}, \quad R^5{}_{151} = \frac{w^2}{l^4 U}, \quad R^1{}_{010} = -\frac{w^2 U}{l^4} - \frac{UU''}{2}, \quad R^1{}_{551} = -\frac{1}{w^2} \quad (3.3a)$$

$$R^0{}_{110} = -\frac{U''}{2U} - \frac{w^2}{l^4 U}, \quad R^0{}_{550} = -\frac{1}{w^2}. \quad (3.3b)$$

Note that the components from the 2 – nd row can in fact be obtained by permutations of the indices 0, 1, 5 from the components of the first row.

Keeping in mind that $\xi^\alpha u_\alpha = 0$ and $u^\alpha u_\alpha = -1$, we arrive, after some calculations, at

$$\frac{d^2 \xi^5}{d\lambda^2} + \frac{1}{l^2} \xi^5 = 0, \quad (3.4)$$

with the solution

$$\xi^5(\lambda) = \xi^5_{(0)} \sin \frac{\lambda}{l}, \quad (3.5)$$

The amplitude $\xi^5_{(0)}$ is a constant of integration. It means the separation on the fifth direction between two geodesic observers oscillates harmonically with the frequency $l^{-1}$. A similar dependence was obtained for $w(\lambda)$.

After some manipulations, we have, for $\xi^0(\lambda)$ and $\xi^1(\lambda)$

$$\frac{d^2\xi^0}{d\lambda^2} + \frac{1}{l^2}\xi^0 = -\frac{U''}{2U}\left(\frac{l^2 w_0^2 U}{w^4}\xi^0 + \frac{El^5}{w^5}\sqrt{\frac{w_0^2}{w^2}-1}\,\xi^5\right) \qquad (3.6)$$

$$\frac{d^2\xi^1}{d\lambda^2} + \frac{1}{l^2}\xi^1 = -\frac{U''}{2}\left(\frac{l^2 w_0^2}{w^4}\xi^1 + \frac{l^5}{w^5}\sqrt{\left(E^2 - \frac{w_0^2 U}{l^2}\right)\left(\frac{w_0^2}{w^2}-1\right)}\,\xi^5\right). \qquad (3.7)$$

Having known the expression (3.5) for $\xi^5(\lambda)$ and (2.9) for w($\lambda$), the functions $\xi^0(\lambda)$ and $\xi^1(\lambda)$ can be obtained once the differential equations (3.6) and (3.7) are resolved.

It is worth to observe that with U(r) a linear function, the r.h.s. of the previous two equations vanish and the dependence of $\xi^0$ and $\xi^1$ on $\lambda$ is the same as for $\xi^5$ (on the contrary, $\xi^5$ is always independent on U(r)). We should have a linear U(r) near the horizon [9] r = 2m, when U (r) ~(r-2m)/2m. In the case of a Minkowskian 4- dimensional spacetime on the brane ( U = 1), the dependence of the components of the separation vector $\xi^\alpha$ on the affine parameter $\lambda$ is the same : $\xi^0$, $\xi^1$ and $\xi^5$ oscillates harmonically with the same frequency $l^{-1}$.

### 4. Conclusions

In this paper we have studied geodesics and geodesic deviation phenomenon in the Chamblin et al. black string geometry , using different coordinates. An inversion transformation z $\rightarrow$ $l^2$/z $\equiv$ w on the fifth coordinate leads to a very simple form of the geodesics. In terms of the affine parameter $\lambda$ along a timelike geodesic, the new coordinate w($\lambda$) oscillates harmonically with the frequency $l^{-1}$ while the radial equation r*(t) is the same as that for a Schwarzschild test particle (r* is the tortoise coordinate). w($\lambda$) is a straightline for a null geodesic and the null radial geodesic has the same mathematical expression as r*(t).

Once the components of the Riemann tensor are computed in the metric (2.5), it is found that the fifth component $\xi^5(\lambda)$ of the separation vector between two nearby geodesics with $\theta = \pi/2$, L = 0 oscillates with the frequency $l^{-1}$, independent of the Schwarzschild potential U(r). Even if the behaviour of $\xi^0(\lambda)$ and $\xi^1(\lambda)$ depends on U, they undergo harmonic oscillations when U(r) is a linear function. We shall study in more detail the eqs. for $\xi^0(\lambda)$ and $\xi^1(\lambda)$ in a future paper.

### Acknowledgements

The author would like to thank Emilian Dudas and Marco Cavaglia for many stimulating and useful comments on the manuscript and one of the referee of CQG for helpful comments.